\documentclass[10pt,twocolumn,showpacs,amsmath,amssymb,floatfix,aps,pre,sort&compress]{revtex4-1}

\usepackage{graphicx}
\usepackage{dcolumn}
\usepackage{bm}

\usepackage{verbatim}

\begin{document}
\title{Frustration of the isotropic-columnar phase transition of colloidal hard platelets by a transient cubatic phase}
\author{Matthieu Marechal}
\altaffiliation{Currently at 
HHU, 
D\"usseldorf, Germany}
\author{Alessandro Patti}
\altaffiliation{Currently at 
IQAC-CSIC,
Barcelona, Spain}
\author{Matthew Dennison}
\author{ Marjolein Dijkstra}
\affiliation{Soft Condensed Matter, Debye Institute for NanoMaterials Science, %
Utrecht University, Princetonplein  5, 3561 RT Utrecht, %
The Netherlands}



\date{\today}

\begin{abstract}
Using simulations and theory, we show that the cubatic phase is metastable for three model hard platelets. The locally favored structures of
perpendicular particle stacks in the fluid prevent the formation of the columnar phase through geometric frustration resulting in vitrification.
Also, we find
a direct link between structure and dynamic heterogeneities in the cooperative rotation of particle stacks, which is crucial for the
devitrification process. Finally, we show that the life time of the glassy cubatic phase can be tuned by surprisingly small differences
in particle shape.
\end{abstract}

\maketitle

Nucleation is the process whereby a metastable phase transforms into a stable one, via the spontaneous formation of a cluster of the stable phase.  According to classical nucleation theory,  the free-energy barrier that separates the metastable phase from the stable state decreases with increasing supersaturation, and for quenches  in  the spinodal regime, the phase transformation  proceeds via spinodal decomposition and coarsening. However, at sufficiently  high supersaturations the motion of the particles can slow down so dramatically  that the metastable state enters a glass regime.

Vitrification  hampers the phase transformation as the particles cannot rearrange diffusively to form the stable phase. However, some glasses can evolve into the stable phase despite the arrested motion. The mechanism behind this so-called "devitrification" process is not well-understood, and neither the origin of the glass transition and its interplay with nucleation. An intriguing scenario based on geometrical frustration has been proposed, where the local order in the liquid phase is incompatible with  the long-range order  of the crystal phase \cite{shintani2006}. Hence, the formation of locally favored structures in the liquid, a concept  proposed by Frank to explain dynamic arrest in glassy systems \cite{frank1952},  prevents the crystallization. This scenario has been investigated using a two-dimensional lattice-free spin glass model, where the degree of frustration against crystallization can be tuned by an additional anisotropic potential that locally favors five-fold symmetry which is incompatible with the crystalline ground state of this model \cite{shintani2006}.

\begin{figure}[b!]
\hspace*{-1em}\includegraphics[width=0.44\textwidth]{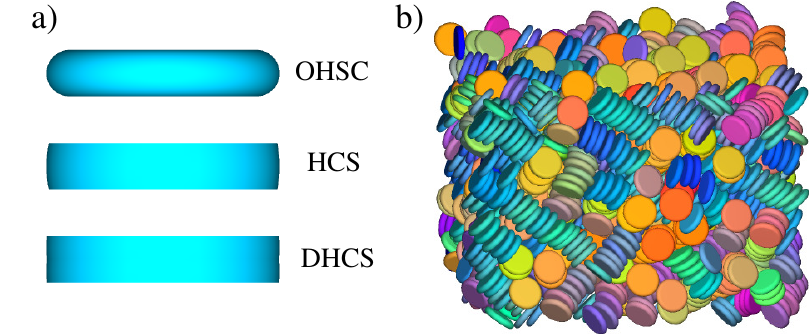}
\caption{(Color online) (a) Three model platelets: oblate hard
spherocylinders (OHSC), hard cut spheres (HCS) and ``double hard cut spheres'' (DHCS). The  volumes of the particles are given by
$v_\text{\tiny OHSC} = \pi L^3/6  + \pi^2 \sigma L^2/8  + \pi \sigma^2 L
/4$ with $\sigma = D-L$, $v_\text{\tiny HCS} = \pi L (3 D^2 -L^2)/12 $,
and $v_\text{\tiny DHCS} = \pi L (3 D^2 - (L/2)^2)/12 $ with $L$
and $D$ the (total) thickness and diameter of the particles,
respectively. (b) A typical configuration of a cubatic phase of
OHSC with $L/D=0.2$ and $P^*=11.25$ ($\eta\simeq 0.57$). Different colors denote different orientations.
\label{fig:shapes}}
\end{figure}

In this Letter, we investigate the interplay between nucleation, geometrical frustration, and devitrification in a simple (more realistic) 3D model system of
colloidal hard platelets using computer simulations. The most common model systems for colloidal platelets are hard cut spheres (HCS), which consist of the middle
section of thickness $L$ of a sphere of diameter $D$, and oblate hard spherocylinders  (OHSC), comprising of a flat cylindrical core with diameter $D$ and height
$L$, and a toroidal rim with tube diameter $L$.  OHSC are more rounded than HCS (see Fig. \ref{fig:shapes}a), and are therefore expected to better model the
shape of colloidal disks, such as polymer-coated clay
platelets~\cite{Qazi2010} or charge-stabilised Gibbsite platelets~\cite{Mourad2009}. Interestingly, for $L/D = 0.2$, the phase diagram of OHSC displays an
isotropic-columnar (IC) phase transition \cite{MM_Cuetos2011platelets}, whereas a very  peculiar  {\em cubatic} phase was reported in between the isotropic and the
columnar phase for HCS  \cite{Veerman_Frenkel}. In this phase, the particles form small stacks of almost cube-like dimensions, which tend to align perpendicular to each other. Recently, it was shown that larger  system sizes tend to destabilize the cubatic
phase~\cite{duncan_cut_spheres,duncan2011}.  However, it remains an open question whether or not the cubatic phase  is  thermodynamically stable for colloidal hard platelets.

Here, we show that the  cubatic phase of different model hard platelets is not  stable, but should be considered as a transient phase in the IC phase transformation.  In addition, we show that the degree of geometric frustration can be altered via subtle changes in the particle shape: the life time of the cubatic phase increases considerably for particles with sharper edges.

We first consider a suspension of $N$ OHSC with aspect ratio $L/D=0.2$ in a volume $V$ or at a pressure $P$. This system displays a bulk transition  from an
isotropic phase with packing fraction $\eta_I \equiv  v_\text{OHSC}N/V=0.5050$ to a columnar phase with $\eta_C =0.5691$ at pressure $P^*=\beta P
v_\text{OHSC}=8.27$, where $\beta=1/k_BT$ and $v_\text{OHSC}$ is the volume of the OHSC particle \cite{MM_Cuetos2011platelets}.

\begin{figure}[t!]
\includegraphics[width=0.48\textwidth]{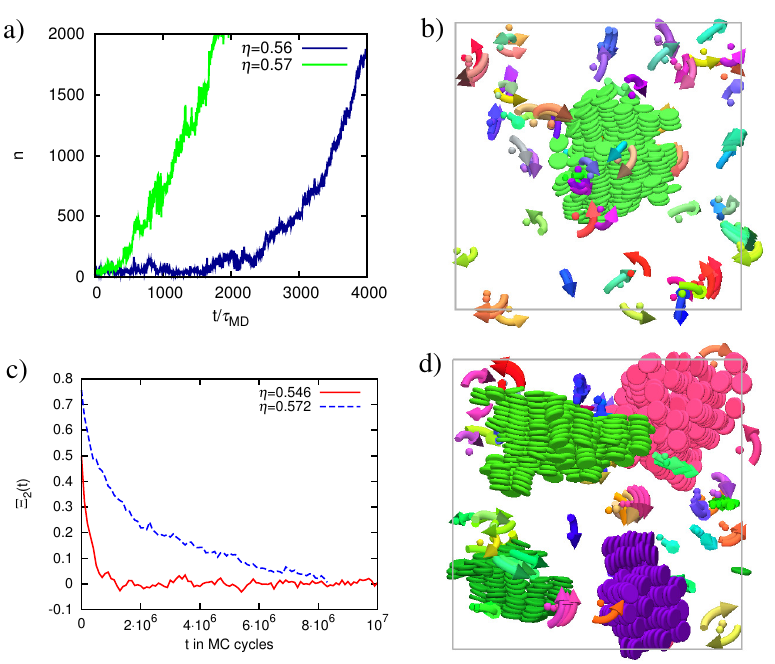}
\caption{(Color online)
(a) The size  of the largest columnar cluster $n$ in an MD simulation of an isotropic fluid  of $N=10000$
OHSC with $L/D=0.2$ at packing fraction $\eta=0.56$ and
$\eta=0.57$. (b)
and (d) Typical configurations of the largest columnar clusters,
when the size  of the largest cluster is around 300 particles
(post-critical) for $\eta=0.56$
(b) and $\eta=0.57$ (d). Different colors of the particles denote
different orientations. Swiftly rotating fluid particles 
are denoted by arrows pointing into
the direction of the rotation; the other fluid particles are not shown at all.  (c) The orientational correlation
function $\Xi_2(t)$ as a function of MC cycles for
$\eta=0.546$ and 0.572.
\label{fig:clust}}
\end{figure}

In order to study the spontaneous formation of the columnar phase from the isotropic fluid phase, we require a cluster criterion that enables us to identify the columnar clusters. Unfortunately, the cluster criterion that was introduced  to study nucleation of the nematic, smectic, and crystal phase in systems of colloidal hard  rods \cite{Cuetos2007,patti2009,Ni2010prl}, is not strict enough to
identify columnar clusters.  We therefore developed a new cluster criterion that enables us to detect columnar clusters.
Particles are considered eligible for inclusion into a columnar cluster 
if they have sufficient neighbors with columnar order.
Neighbors are considered to have columnar order when they have sufficient hexagonal order, as measured
by a standard order parameter, but do not show high ordering with another symmetry.
See~\cite{suppinfo} for details.

\begin{figure}[t!]
\includegraphics[width=0.40\textwidth]{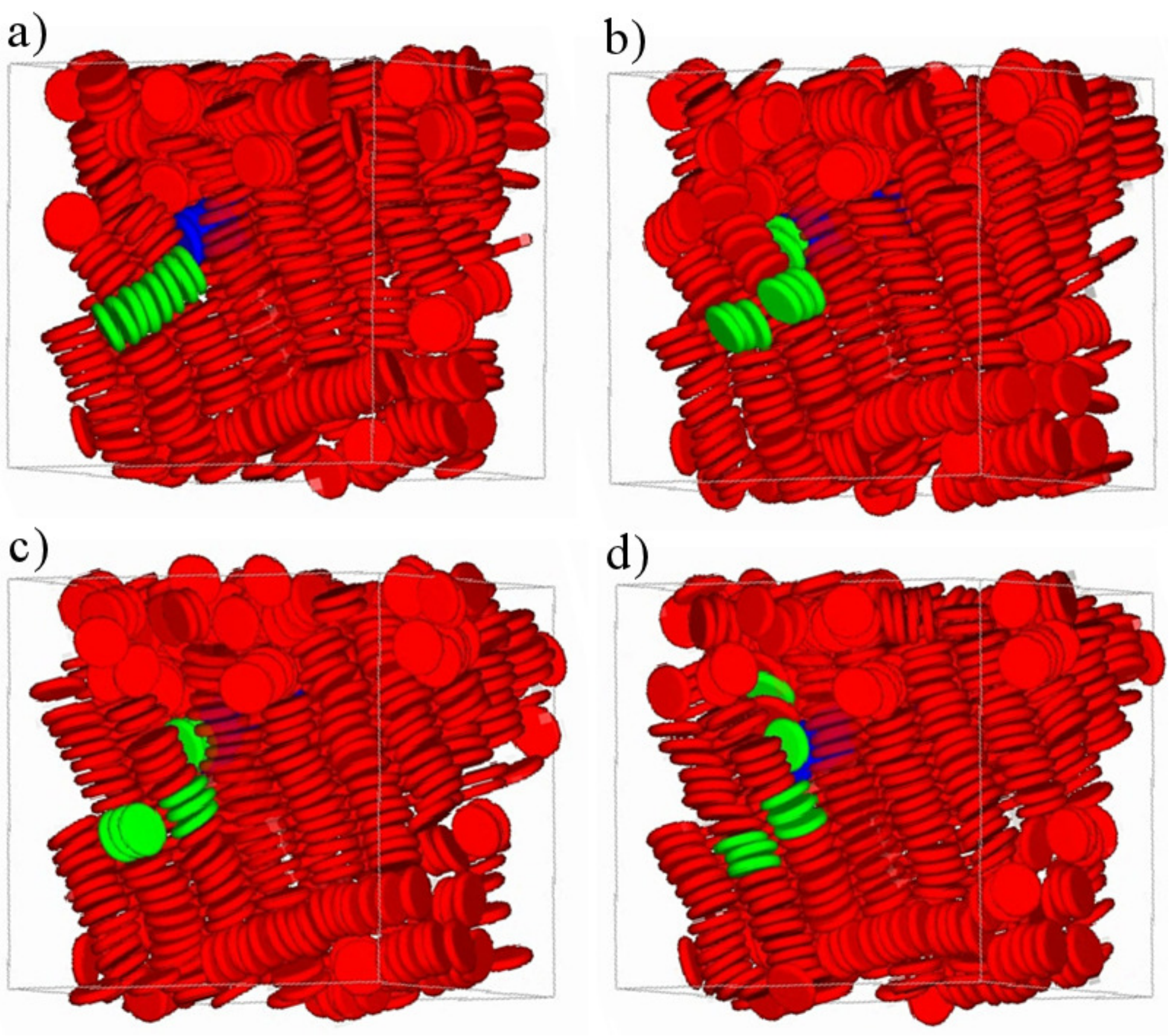}
\caption{(a) A long stack
of (green) particles, which is oriented perpendicular to the nematic director of the columnar phase, is divided in smaller packages of two to four particles as shown in (b). Subsequently, these smaller
packages of particles can rotate by 90 degrees as shown in (c) and (d) in such a way that the orientation matches the
columnar phase.
\label{fig:rot}}
\end{figure}

We use event driven Molecular Dynamics (MD) simulations of relatively large
system sizes ($N=1500, 3000,$ and $10000$) to study the kinetics of the IC phase
transformation. Time is measured in units of $\tau_\text{MD}=\sqrt{\beta m D^2}$, where $m$ is the mass of an OHSC. For $\eta \geq 0.56$, we observe the immediate formation
of short stacks of OHSC in the supersaturated isotropic fluid phase, which
subsequently tend to orient perpendicular to each other to optimize the  packing.
The symmetry  of these locally favored structures in the resulting cubatic phase is incompatible with that of the
columnar phase, and hence the cubatic orientational order can be seen as geometric
frustration against the formation of the columnar phase. The cubatic order is more
pronounced and longer ranged for higher $\eta$ and smaller system sizes ($N=1500$ and 3000). A typical configuration of such a cubatic phase is shown
in Fig. 1(b) for $\eta=0.57$. However, in very long simulations the cubatic phase always transforms into a columnar phase.
In order to analyze the phase transformation, we show  the time evolution of the largest columnar cluster identified by  our cluster criterion  in Fig.~\ref{fig:clust}a for $\eta=0.56$ and 0.57. We clearly observe that the
cluster grows much slower for lower $\eta$ due to the lower supersaturation. In addition, we present typical
configurations of post-critical columnar clusters in Fig.~\ref{fig:clust}b and d.
At a packing fraction of $\eta=0.56$, we observe one columnar cluster that grows further to
form the stable columnar phase, while for $\eta=0.57$ three post-critical clusters are observed
as the nucleation barrier is much lower. Interestingly, the nematic directors of the columnar clusters are aligned along the three preferred axes of the cubatic phase, where it originated
from, see Fig.~\ref{fig:clust}d. We conclude that the IC phase transformation  proceeds via a transient cubatic phase and corresponds to a nucleation and growth scenario in which a spontaneously formed columnar cluster grows out to form the stable columnar phase.
During  MD simulations with $N=10000$, we observe the appearance of columnar clusters
before long-range cubatic order appeared after quick compression.
In fact, this was to be expected, as the time it takes for the cubatic
order to spread throughout the system increases with system size, while the time
for a nucleus to form in a  fixed volume is system size independent.
Therefore, the behavior observed in experiments, such as the Cryo-TEM
experiments in which the
cubatic phase was observed~\cite{Qazi2010}, is largely dependent on the
sample volume,  which for Cryo-TEM is rather small to allow sufficiently fast
shock-freezing of the sample.
This suggest that  the cubatic phase may be stabilized by confinement.

Additionally, heterogeneous dynamics in the form of collective particle reorientations is
found in the locally cubatic fluid, as can be appreciated from Fig.~\ref{fig:clust}b and d, 
where we represent all fluid particles that rotate swiftly by arrows denoting the direction of the rotation.
We wish to remark here that the clusters of particles which rotate simultaneously
can be easily identified here as small stacks of up to four particles,
which is impossible for glassy states of spherical particles where heterogeneous dynamics
cannot  be easily related to the local structure~\cite{Sanz_cryst_glass}.
Similarly, the growth of a columnar phase
often proceeds by collective attachment
of small stacks rather than single particles~\cite{suppinfo}. Interestingly, the rotation of stacks 
also plays a crucial role in the late-stage development of the columnar cluster.
Fig.~\ref{fig:rot} shows a form of defect-healing in which a stack of mis-aligned particles in the columnar cluster first breaks up into
smaller stacks and, subsequently, these smaller stacks reorient to conform with the director of the cluster.

As nucleation of a columnar phase from a glassy state with cubatic order is hardly
studied, it is interesting to determine the nucleation barrier associated for this devitrification process.
Since the equilibration can only proceed via  collective rearrangement of small clusters, the formation of the columnar phase is severely hampered by slow dynamics. We determine the nucleation barrier by employing the umbrella sampling technique in Monte Carlo (MC) simulations \cite{Cuetos2007}.
To obtain an estimate of the number of MC cycles needed to sample
the configurational space sufficiently, we  study first  the
translational and orientational dynamics of the particles in the
cubatic phase. To this end, we calculate the mean square
displacement  $\langle (\mathbf{r}_i(t)-\mathbf{r}_i(0))^2\rangle$
(not shown) and the orientational correlation function $\Xi_2(t)
= \left \langle \tfrac{3}{2} \lbrack{\bf u}_i(t)\cdot  {\bf u}_i(0)\rbrack^2-\tfrac{1}{2} \right \rangle $, where the angular brackets indicate
an ensemble average and $t$ indicates the  time measured in MC cycles.
Exemplarily, Fig. 2b displays $\Xi_2(t)$  for
$\eta=0.546$ and 0.572. Similar behavior is seem in the event-driven MD simulations~\cite{suppinfo}, where a clear slowing
down of the translational and rotational dynamics by about a
factor of $\simeq 6$ is observed when the packing fraction is increased from
$\eta=0.5$ to $0.56$.
The simultaneous slowing down of  the translational and rotational dynamics should be contrasted with the  decoupling of the freezing of the translational and rotational degrees of freedom that is found for
ellipsoids~\cite{DeMichele2007ellipsoids,Letz2000}.

\begin{figure}
\includegraphics[width=0.37\textwidth]{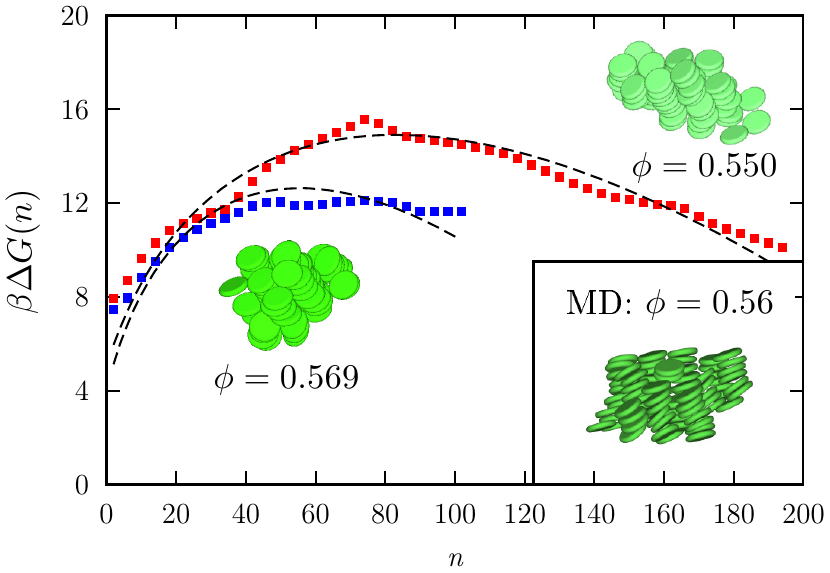}

\caption{
Gibbs free energy $\beta \Delta G(n)$ as a function of the number of particles $n$
in the largest columnar cluster for a system of OHSC with
$L/D=0.2$ and packing fraction $\eta=0.550$
and $0.569$  (squares). The dashed line is a fit~\cite{Filion2010hsnucl} to classical nucleation theory.
Configurations of the critical nucleus  are shown obtained from MC
for $\eta=0.550$  and $0.569$
 in the main
panel
 and from MD simulations at
$\eta=0.56$  in the inset.
\label{fig:nucl}}
\end{figure}

Subsequently, we determine the Gibbs free energy $\Delta G(n)$ as a function of columnar cluster size $n$ using the umbrella sampling technique \cite{Cuetos2007}.
The nucleation barriers for OHSC with $L/D=0.2$ and  $\eta=0.550$ and $0.569$ are shown in Fig.~\ref{fig:nucl}.
As equilibration is rather
slow due to glassy dynamics, the noise on  $\Delta G(n)$ is significant. However, we are able to determine for the first time a nucleation barrier for a devitrification process in which a glassy state transforms into a stable phase via  collective particle reorientations.
We also present typical configurations of the critical nucleus
that corresponds with the top of the barrier in
Fig.~\ref{fig:nucl} along with the critical nucleus obtained from
MD simulations at $\eta=0.56$. We find that the results for the
structure and shape of the critical nucleus as obtained from MC or
MD simulations are very similar, i.e., the cluster consists of an
hexagonal array of particle stacks and the overall shape of the
cluster is roughly spherical. The barrier height  $\beta \Delta
G^*(n)\simeq 15$ and 12 for $\eta=0.550$ and 0.569, respectively.
We mention here that the  corresponding values for the
supersaturation $\beta \Delta \mu=0.414$ and 0.553, which is  the
driving force for nucleation, is extremely small, for comparison
we note that, for hard spheres at $\eta=0.5478$, $\beta \Delta
\mu\simeq 0.74$, which leads to fast nucleation~\cite{Filion2010hsnucl}.

\begin{figure}[t!!]
\includegraphics[width=0.42\textwidth]{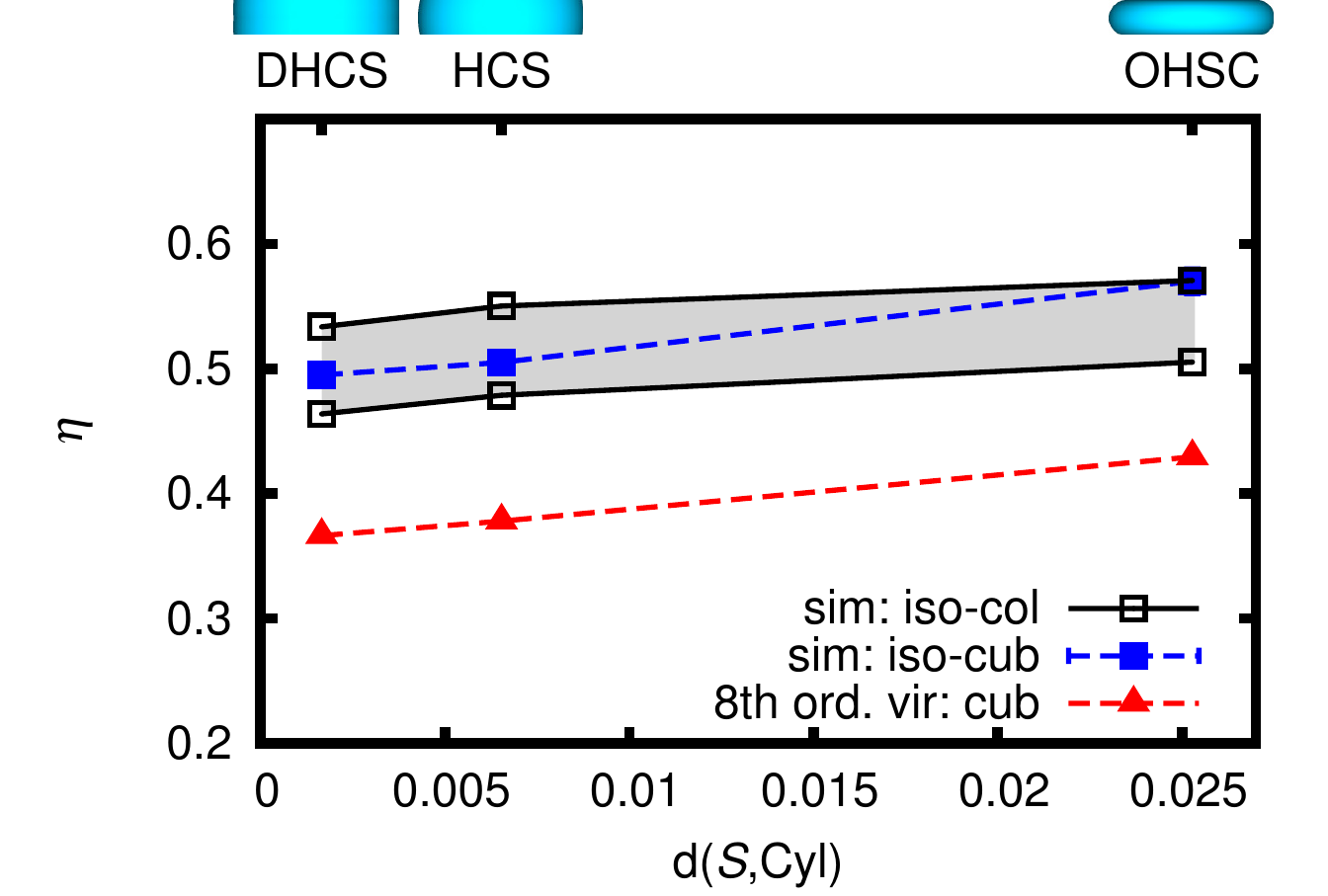}

\caption{
The phase diagram for double hard cut spheres (DHCS), hard cut spheres (HCS) and OHSC with $L/D=0.2$: 
packing fraction $\eta$ versus $d(S,\mathrm{Cyl})$, the difference~\cite{suppinfo,moszynska2006convex}
between the shape $S$ in question and a cylinder (Cyl) with the 
same volume and aspect ratio.
The gray area and the black lines denote  coexistence between the isotropic  and the columnar phase from simulations~\cite{suppinfo}.
The metastable transition from the isotropic to the cubatic phase is denoted by
the dashed lines, where the solid squares denote simulation results for
$N=3000$ particles, and the triangles denote the results from 8th order virial theory. 
\label{fig:phbeh}}
\end{figure}

Our results show clearly that the cubatic phase of OHSC  is metastable with respect to an IC phase transition. In this light, it is interesting to  study the effect of particle shape on the stability of the cubatic phase. To this end, 
we  measure the cubatic order parameter~\cite{duncan_cut_spheres} as a function of pressure for the three particle shapes
depicted in
Fig.~\ref{fig:shapes} using $NPT$ MC simulations
with $N=3000$ particles.
The shapes (ordered from more curved to more
cylinder-like) are OHSC, HCS, which resemble recently synthesized particles of Ref.~\cite{Fujibayashi2007},
and the double hard cut spheres (DHCS). The latter model consists of two superimposed
 HCS and is essentially a cylinder. All three models have the same height-to-diameter ratio of $L/D=0.2$.
The onset of cubatic order, as defined by the packing fraction at
which the cubatic order parameter suddenly increases,
is shown as the dashed, blue line in Fig.~\ref{fig:phbeh}. We observe clearly that the cubatic order sets in at lower $\eta$ upon decreasing the particle curvature.

In addition, we study the stability of the cubatic phase using a high-order virial theory. In  Ref.~\cite{duncan_cut_spheres}, it was shown that a 8th-order
expansion is required to predict a stable isotropic-cubatic  phase transition for HCS with  $L/D=0.2$. Here, we apply this theory  to predict the
isotropic-cubatic phase transition for  OHSC, HCS (with higher precision than in Ref. ~\cite{duncan_cut_spheres}), and DHCS.
The results for the bulk density of the cubatic phase in coexistence with the isotropic phase are denoted by the dashed, red lines in
Fig.~\ref{fig:phbeh}c at the 8th virial level. Although, the agreeement with the simulations results is not satisfactory, surprisingly, the trend of the bulk density of the cubatic phase with particle curvature is very similar to that of the $NPT$ MC simulations.

Finally, we determine the stability of the cubatic phase with respect
to the columnar phase for HCS for which this phase was originally
discovered~\cite{Veerman_Frenkel}.  The apparent stability of the cubatic phase for HCS could be due to dynamic arrest. Inspired by the particle stack rotations as observed in our MD simulations, we introduce a new  cluster move in the MC simulations to speed up equilibration \cite{suppinfo}.
In order to investigate more precisely whether or not there is phase coexistence
between an isotropic fluid phase with either a cubatic phase or a columnar phase, we perform $NPT$ MC simulations of the two coexisting phases in a  simulation
box that is large enough that the interfacial free energy is sufficiently small \cite{vega,Zykova-Timan_coex_HS_AO}.  The phase that grows at the expense of the other phase must be the
stable phase. 
The pressure at coexistence can be determined as the pressure at which the growth speed of the columnar phase is zero.
The corresponding coexistence densities for HCS and DHCS are shown in Fig.~\ref{fig:phbeh}. Clearly, the pressure at which the columnar phase
becomes more stable than the isotropic phase is lower than the pressure at which long-range cubatic order was found,
which unambiguously shows that the cubatic phase is unstable for all three shapes considered.

In conclusion, we find that the cubatic phase is metastable with respect to an IC phase coexistence for all three model
platelets, and can be regarded as a transient phase in the IC phase transformation. The locally favored structures of perpendicularly oriented particle stacks in the cubatic phase leads to geometric frustration that prevents the formation of the columnar phase thereby yielding vitrification. Additionally, we find a direct link between  structural order  and  dynamic heterogeneities provided by the cooperative rotation of particle stacks in the cubatic phase. Such a link is often assumed to be characteristic for glassy behavior, but is not easy to demonstrate in e.g., colloidal hard sphere glasses. We also show that the cooperative stack rotations play an important role in the devitrification process and that the life time of the cubatic phase can be tuned by confinement and by surprisingly small differences in the particle shape. Interestingly, our results explain recent experimental observations on suspensions of  gibbsite platelets which enter a kinetically arrested glass regime upon increasing the particle concentration and in which  small iridescent grains of the columnar phase were formed after periods of months to years \cite{Mourad2009}.

\end{document}